# Intertrochanteric Fracture Visualization and Analysis Using a Map Projection Technique

Yucheng Fu[a†], Rong Liu[ab†], Yang Liu[a*], Jiawei Lu[c]

[a]Nuclear Engineering Program, Mechanical Engineering Department, Virginia Tech
635 Prices Fork Road, Blacksburg, VA 24061
[b]Department of Orthopadics, PuRen Hospital Affiliated with Wuhan University of Science and Technology
No.1 Benxi Road, Qingshan District, Hubei Province, China 430080
[c]Department of Orthopadics, First Affiliated Hospital, Dalian Medical University
Dalian, China 116044
ycfu@vt.edu; dr_liurong@whu.edu.cn; liu130@vt.edu; jwlu307@gmail.com

Total number of words of the manuscript: 5581
Abstract word number: 185
Figure number: 10
Table number: 3

## Abstract

Understanding intertrochanteric fracture distribution is an important topic in orthopadics due to its high morbidity and mortality. The intertrochanteric fracture can contain high dimensional information including complicated 3D fracture lines, which often make it difficult to visualize or to obtain valuable statistics for clinical diagnosis and prognosis applications. This paper proposed a map projection technique to map the high dimensional information into a 2D parametric space. This method can preserve the 3D proximal femur surface and structure while visualizing the entire fracture line with a single plot/view. Using this method and a standardization technique, a total of 100 patients with different ages and genders are studied based on the original radiographs acquired by CT scan. The comparison shows that the proposed map projection representation is more efficient and rich in information visualization than the conventional heat map technique. Using the proposed method, a fracture probability can be obtained at any location in the 2D parametric space, from which the most probable fracture region can be accurately identified. The study shows that age and gender have significant influences on intertrochanteric fracture frequency and fracture line distribution.

[†] Yucheng Fu and Rong Liu contributed equally to the work

# 1 Background

Intertrochanteric fracture (IT) is a common severe injury among seniors that has received much attention due to its high morbidity and mortality. It requires great effort for orthopadic surgeons to provide successful operation. In the past, different systems have been developed for the classification of proximal femur fractures aiming to achieve better diagnoses and treatments [1–4]. However, in these systems, the statistical information of common fracture pattern, high-risk region, etc. are not used as direct criteria for fracture classification. Further, the understanding of the common fracture patterns is of great importance to help surgeons to make proper open reduction plans and internal fixation strategies for those that need surgery.

The proximal femur fracture can be largely classified into three main categories based on the anatomical position of the fracture line, namely, femoral neck fracture, intertrochanteric fracture, and subtrochanteric fracture. Among these type of fractures, IT fracture constitutes nearly half of the hip fractures caused by low energy trauma [5] and is particularly common in seniors. From 1999 to 2012, the IT fracture still yields a high mean one-year mortality rate of 23% [6]. Therefore, this study focuses on the visualization and analysis of IT fracture to obtain a better understanding of its distribution and mechanism. The IT fracture is referred to the fractures in the region, which lies below the proximal femur neck and above the bottom transverse plane of the lesser trochanter.

In the past ten years, the fracture map technique has been developed to visualize fracture patterns and obtain statistical characteristics from large clinical datasets [7]. One can map individual fracture lines to a standard bone template for visualization and analysis. With enough samples, the fracture line direction, pattern and frequency information can be visualized directly on the standard template. The fracture map can provide accurate and consistent information about the actual fracture morphology since it is based on statistical results from clinical data. It can be used to develop a new classification system or sub-classifications of existed classification systems based on real fracture patterns and comminution zones [8]. Fracture prognosis prediction, hardware placement, and implant design optimization can also benefit from the use of fracture map [8–10]. In the past decade, the fracture heat map technique has been applied to scapular fracture [11], tibial pilon fracture [7], radial head fracture [12], and tibial plateau fracture [8], etc.

The procedures of using the fracture heat map technique can be summarized in four steps. The first step is to reconstruct and reduce the 3D mesh/model of specific bone from CT scan data. The second step is to identify the fracture line or fracture region from the reconstructed model. The third step is to map all fracture lines or regions acquired from different patients to a standard template. Fracture pattern or heat map can be visualized when all clinical data are mapped to the standard template. The fourth step is to evaluate the 3D model which presents the fracture line or heat map and to select a proper tomography view which can project and visualize the fracture pattern.

The selection of the view usually depends on the nature of the fracture patterns. If a fracture line is across several different anatomical positions, two or more views are required to describe the fracture pattern [8] fully. With the consideration of different factors, such as age and gender, the visualization and analysis can become complicated and cumbersome with too many fracture maps at a time. In addition, the scaling distortion exists when a 3D model is directly projected onto a 2D plane. In this case, the length, affected area and spatial distribution of fracture lines cannot be accurately represented in the 2D projected plane. Statistical information obtained from these fracture maps may be erroneous due to distorted dimension scales. To address these issues, this paper proposes a new method to present the fracture map based on the map projection technique [13,14]. The proposed method unfolds a 3D bone mesh and maps it into a 2D parametric space, which retains all the geometrical and topological information of the fracture. The scaling factors are all kept in the mapping function so that accurate statistical information can be obtained from the 2D parametric maps even with distortion. This paper shows that the proposed fracture map technique can display the complete intertrochanteric fracture line in one image, which may otherwise need all four views (anterior, medial, posterior and lateral) using conventional techniques. The new technique can significantly facilitate the visualization and analysis of statistical information obtained from a large number of patients. The influence of gender and age on the IT fracture line distribution are studied in this paper based on the proposed technique.

## 2  Methods

2.1  Subjects

A total of 100 IT fracture cases are used in this study. The data are collected from PuRen Hospital (Wuhan, China) over an approximately three-year period from December 2013 to January 2017. This study was approved by the PuRen Hospital Institutional Ethics Committee (IEC). The selected cases contain preoperative CT scan data, and the fracture region is confined to the IT region which is the interest of this study. All the data are acquired by Siemens SOMATOM Sensation 16 CT. The CT scan images with a slice thickness of 1.5 mm or below are included to ensure the data quality.

A standard proximal femur template is used in this study for IT fracture visualization. The standard template is acquired from CT data of a normal right proximal femur. The proximal femur region is extracted by applying a threshold of 226 Hounsfield units (HU) to the original CT image. The region extracted in each CT slice are then combined to generate a standard 3D model. The model is then processed to generate the stereolithography (STL) file which will be used in the following steps for superimposing the fracture lines of different patients.

2.2  Fracture region labeling and heatmap generation

For each patient case, the proximal femur is reconstructed from 2D CT images, and the fracture segments are reduced to the anatomical position if displacement happened. A rigid registration step is followed to align the reconstructed proximal femur to the standard template by rotation and translation. Five anatomic landmarks are chosen for the rigid registration task. As shown in Fig. 1 (a), the five landmarks include the mass center of the femur head (point 1), the left (point 2), and right (point 3) greater trochanter apex in lateral view, the apex of lesser trochanter (point 4) and the intersection point (point 5) between the anatomical axis and the transverse plane passing through the lesser trochanter apex. These five landmarks are important locations in fracture diagnosis and prognosis, as well as in surgical procedures. With proper alignment, a trained technician drew the fracture lines of each case on the standard model by identifying the fracture lines on the patient's proximal femur model. All fracture lines drew on the standard template are independently reviewed by an orthopaedic surgeon to ensure accuracy and quality. An example of the fracture lines drawn on the standard template by referring to the patient's 3D reconstructed proximal femur model is shown in Fig. 1 (b). The fracture region is marked by a 4-mm-wide line. The line is wide enough to cover the actual fracture region. Compared to a zero-thickness fracture line, the 4-mm-wide line used here can be mapped directly onto the 2D parametric space or 2D projection views for frequency calculation or heat map visualization. Compared to the fracture line with a zero thickness, this method requires no interpolation on the fracture map for heat map generation. Note that the scaling factor on the fracture map varies by location after mapping onto the parametric plane. Distortions will be introduced if interpolation is made to the fracture lines on the heat map. When fracture cases are superimposed onto the standard template, the heat map can be used for an intuitive representation of fracture counts, the percentage of involvement, etc.

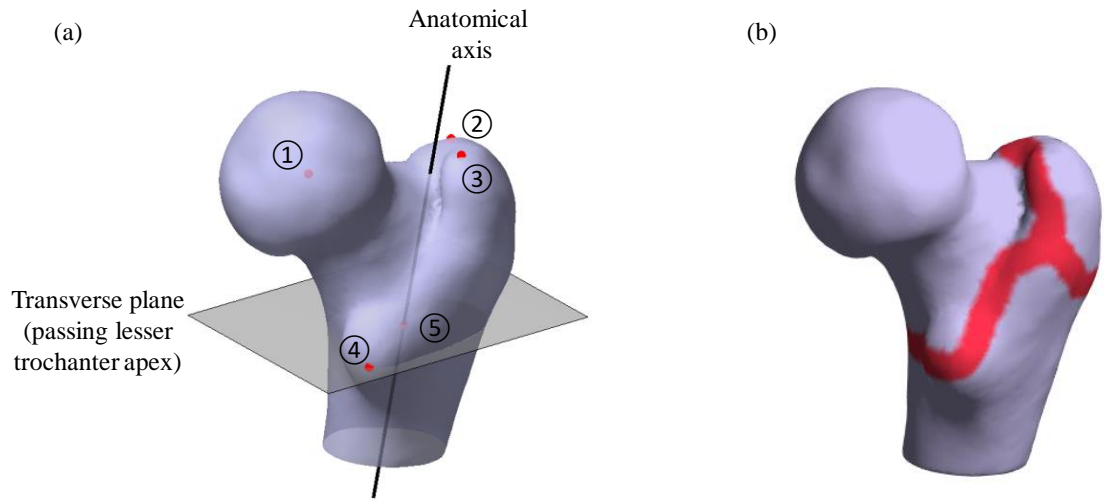

Fig. 1 Procedures of mapping fractures to the standard template (a) The selection of anatomical landmarks for rotation and alignment, (b) an example of a fracture line drawn on the standard template.

2.3   2D parametric fracture map

The fracture line on the bone surface is typically a complicated 3D curve. To study the fracture region, the transformation of 3D information to a 2D plane is a challenging task. Conventional methods project the 3D bone mesh onto a 2D surface or a simplified 2D sketch template. The fracture lines are then superimposed onto the template. In this process, the fracture information could be lost while presenting the fracture lines with only one view. Two or more views are necessary to fully display the 3D information of the fracture line along the bone surface. To present the details of an IT fracture line, four views from different angles are necessary for presentation. To improve the efficiency and accuracy of fracture line visualization, a map projection technique is developed aiming to visualize the complete fracture information of proximal femur in a 2D plane. Similar to cylindrical map projection [15] used in geography, the technique used in this paper unfold the proximal femur to a 2D parametric space.

The details of this mapping are shown in Fig. 2. In the figure, the proximal femur is first plotted in the Cartesian coordinate with axes of $x, y, z$. An unfold line, which is marked by a red dash line, is used for spreading out the 3D surface into the 2D parametric space. To determine the unfold line location, the coronal plane is placed by passing the apex point of the femoral head. The intersection line between the 3D model and the coronal plane are selected as the unfold line. The unfold line starts from the apex point of the femoral head, goes down along the medial side and ends at the bottom of the subtrochanter. This unfold line provides a reference for the 2D parametric space. With the given standard proximal femur template, the cross-section shape at each specific height $z = z_i$ can be acquired. At each slice location $z = z_i$, the coordinate $z'$ in parametric space is set the same as $z$ in the physical space. The second coordinate $s'$ is defined as:

$$s' = s / s_{max}(z'), \tag{1}$$

where $s$ is the curve length along the boundary of the slice, starting from the unfolding line red point along a clockwise direction, and $s_{max}(z')$ is the perimeter of the slice at the height of $z'=z_i$. Using this mapping convention, every surface point in the physical space $(x, y, z)$ is projected onto a unique point $(z', s')$ in the 2D parametric space. The parameter $z'$ shares the same range with $z$ in the physical space, whereas the parameter $s'$ has a range of [0,1].

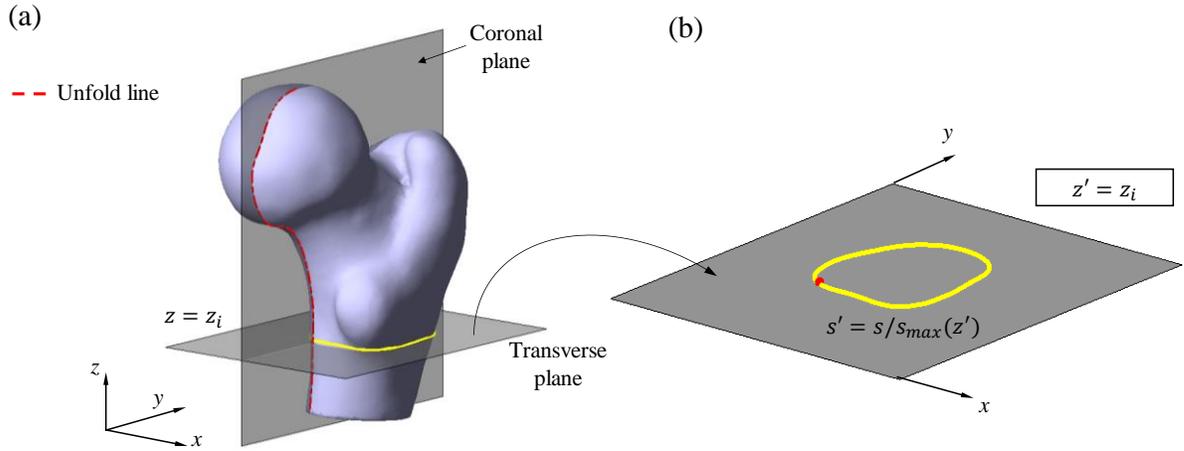

Fig. 2 Pipeline of mapping surface in (a) the physical coordinate to (b) the 2D parametric coordinate using the map projection technique.

For a given surface property $p(x, y, z)$, the distribution of $p$ in the parametric space can be calculated as:

$$p(z', s') = p(z'(x, y, z), s'(x, y, z)) = p(f(x, y, z)), \tag{2}$$

where $f$ is the function mapping the points in physical space $(x, y, z)$ to parametric space $(z', s')$ and the $p$ can represent different parameters on the proximal femur surface. An example is given in Fig. 3. In Fig. 3(a), the proximal femur is divided into six regions on the standard 3D template [1,16]. The six regions include femoral head, neck, greater trochanter, lesser trochanter, intertrochanteric region and subtrochanteric region. For the convenience of discussion, the intertrochanteric region in this study excludes the greater and lesser trochanter as shown in the plot. In this example, the $p$ value should represent the color in each region. By applying the map projection technique, the unfolded surface in the 2D parametric plane is shown in Fig. 3(b). As can be seen in the figure, the projected map includes both the anterior and posterior views of the 3D model. Different regions are completely visible in a single map, and they are intuitive for recognition. In the 2D parametric map, it can be seen that the regions are arranged in a sequence along the $z'$ axis. The relative location of six regions with respect to the peripheral direction is indicated by the $s'$ coordinate. It should be noticed that the femoral head and the greater trochanter area are partially connected in the 2D parametric map. Since the surface is unfolded along the $z$ direction, the slice may contain two separate regions, one of the femoral head and the other the greater trochanter. After being converted into the *s'* coordinate, these two physically separated regions may be in contact with each other in the *z'-s'* space. Therefore, a dotted line is plotted in Fig. 3(b), which indicates that these two regions are separated by a finite distance in the physical space.

Fig. 4 shows a qualitative assessment of the distortion when mapping the 3D surface into the 2D parametric space. The 3D surface was coated with a color gradient texture consisting mainly red, blue and yellow as shown in Fig. 4 (a) and (b). After mapping the 3D surface into the 2D parametric space, the color texture distribution is shown in Fig. 4 (c). Referring to the color distribution on the 3D surface, one can have a visual impression of the distortion during this transformation. As can be seen in the figure, the color gradient in the top femoral head and subtrochanteric region become more diffusive after transformation, which indicates a greater degree of stretch. On the other hand, the greater and lesser trochanter regions show a compressed and striped color patterns. This is because each unit length of *s'* in these regions represents a larger physical distance in the 3D space.

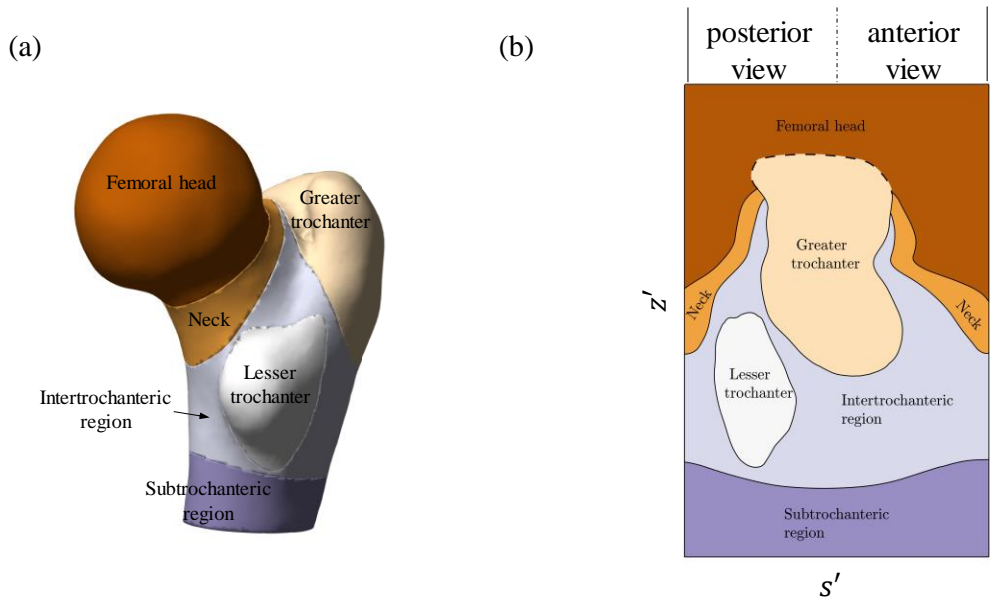

Fig. 3 (a) Proximal femur region classification and (b) 2D parametric map view for all the six labeled regions.

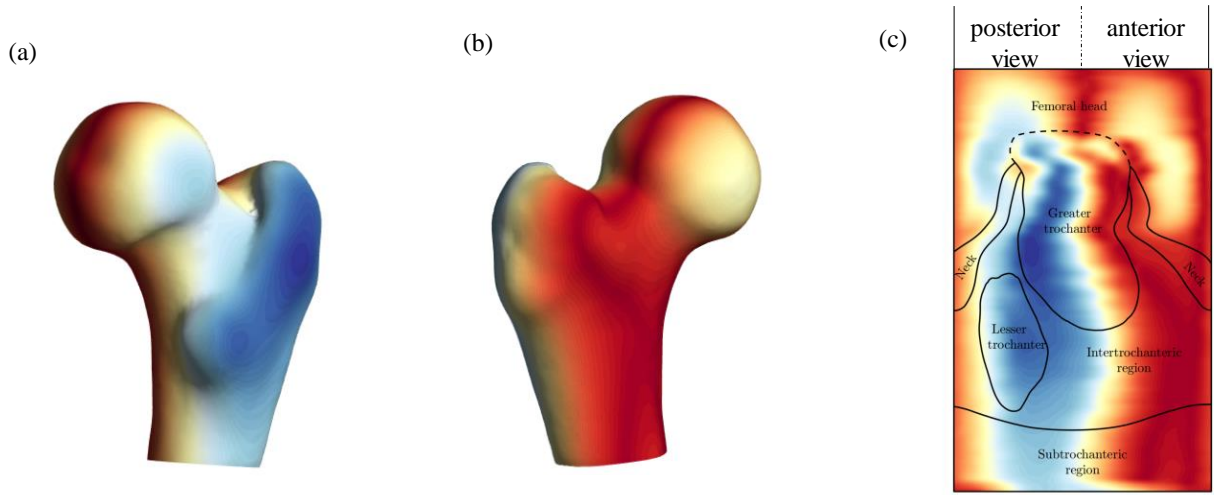

Fig. 4 Visualization of the distortion level when unfolding the 3D surface, which is shown in (a) posterior-medial view and (b) anterior-lateral view into the (c) 2D parametric map. The proximal femur surface is coated with a color texture for visualization.

## 3   Results

The data used in this study consists of 42 male patients and 58 female patients. The average age of all 100 patients is 61 years. The age and gender distributions are plotted in Fig. 5. As shown in the figure, 28 patients fall in the age range of [30, 50) including 9 males and 19 females. Another 39 patients fall in the age range of [50, 70) with 17 males and 22 females. The rest 33 patients have an age over 70 with 16 males and 17 females.

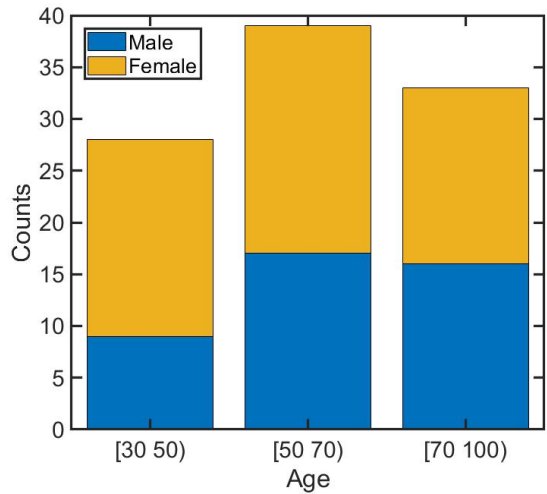

Fig. 5 The age and gender distribution of the 100 patients.

By combining all the 100 cases together, the fracture heat map is first presented in Fig. 6. The color represents the probability of fracture, which is calculated by the number of fracture cases at each location divided by the total number of cases. The conventional four anatomical views: anterior, medial, posterior and lateral, are shown in Fig. 6 (a) to (d). The heat map on the 2D parametric plane is shown in Fig. 6 (e) for comparison. For all the 100 cases, it can be seen in Fig. 6 (e) that the most frequent intertrochanteric fracture region is located in the lower left corner of the lesser trochanter with a probability of around 70%. The high-risk red region passes through the lesser trochanter and extends to the greater trochanter along the intertrochanteric crest on the posterior side of the proximal femur. Passing through the apex of the greater trochanter, another high-risk region gradually develops along the intertrochanteric line from top to bottom on the anterior side. The fracture probability along the intertrochanteric line is around 40%. The distal lateral wall in the greater trochanter and the subtrochanteric region has a scattered fracture distribution and the fracture probability is usually less than 10% in these regions.

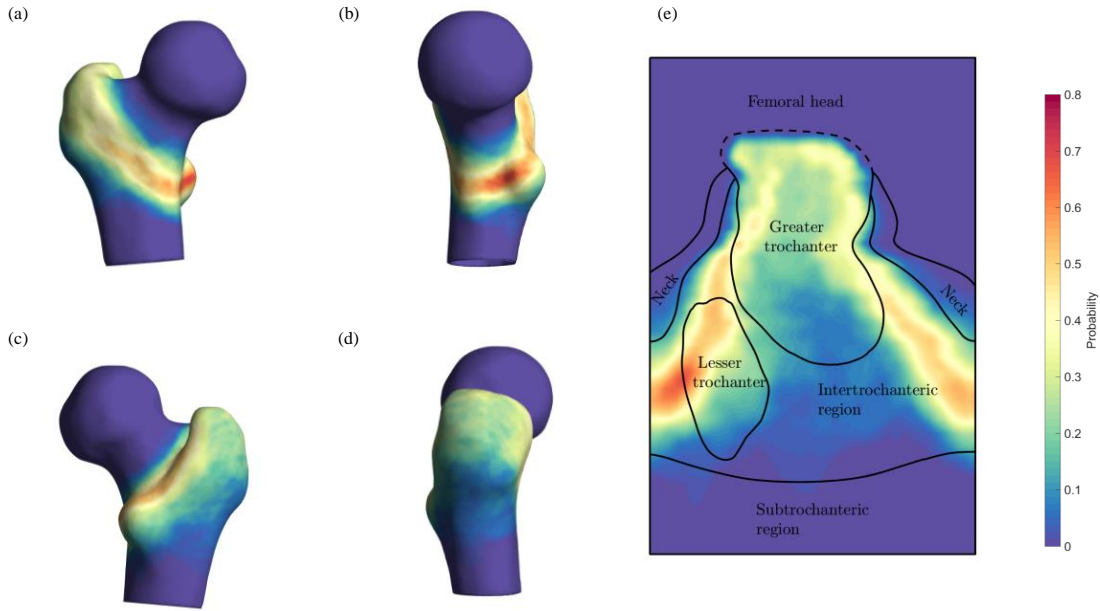

Fig. 6 Proximal femur fracture probability visualization with all 100 cases using four anatomical views: (a) anterior view, (b) medial view, (c) posterior view, (d) lateral view, and (e) the 2D parametric fracture heat map.

The statistical results of these 100 cases are shown in Table 1. The region area shows the surface area of the given region on the standard template. The fracture area in each region are the averaged results calculated from the 100 cases. The number after the plus and minus sign is the standard variance of the fracture area in this region among the studied cases. This indicates the scattering level of the fracture line in each region. The percentage of involvement is the ratio of fracture area to region area. As shown in the table, the greater and lesser trochanter have the largest percentage of fracture area involvement with 24.7% and 29.0%, respectively. The intertrochanteric crest and line occupy the most of the fracture area in the greater and lesser trochanter region. For the intertrochanteric region, it has large fracture area of 940.2±243.8 mm$^2$ but a smaller percentage of involvement 18.2% compared to the greater and lesser trochanter. For the subtrochanteric region, it has a small averaged fracture area of 12.0±34.7 mm$^2$.

Table 1 Averaged fracture area and variance in different proximal femur regions.

|  | Greater trochanter | Lesser trochanter | Intertrochanteric region | Subtrochanteric region |
|---|---|---|---|---|
| **Region area [mm2]** | 4433.3 | 1417.7 | 5179.7 | 2000.6 |
| **Fracture area [mm2]** | 1096.4±474.0 | 410.8±153.6 | 940.2±243.8 | 12.0±34.7 |
| **Percentage of involvement** | 24.7% | 29.0% | 18.2% | 0.6% |

To study the characteristics of IT fracture among different ages, the patients are divided into three age groups as described in Fig. 5. The 2D parametric fracture heat map for the three groups are visualized in Fig. 7. Comparing the fracture risk among three age groups, these 2D parametric fracture heat maps share an overall similar fracture distribution pattern though probability does change locally. One major finding among these plots is the change of the fracture probability in the greater trochanter area. The fracture probability and area of involvement in [50, 70) and [70, 100) groups are larger than the group below 50 in the greater trochanter area. The area with increase fracture probability in greater trochanter is mostly extended from the intertrochanteric crest and intertrochanteric line. The fracture probability distribution in the intertrochanteric region and lesser trochanters are similar among different age groups with smaller fluctuations in probability. The group of 70-year or older shows a slightly spread probability distribution compared to the [30, 50) and [50, 70) groups.

To examine the gender influence on IT fracture distribution, the fracture probability distribution is compared between male and female for the group of 70-year or older. As shown in Fig. 8, the fracture distribution in the female group is more spread out than that in the male. For the female group, the fracture distribution covers nearly the entire greater trochanter and lesser trochanter region, and a major part of the intertrochanteric region. No obvious peaks can be found in the female group. On the contrary, the male group shows a clearly defined high-risk region, similar to but even more concentrated compared to the one shown in Fig. 6 (e) which represents the distribution of all 100 cases.

The user experience of using the proposed 2D parametric map for IT fracture analysis is accessed using a five-level Likert scale table [17] as listed Table 2. Five questions are provided to access the method in different aspects. The evaluation of the 2D parametric map method is based on the three plots shown in Fig. 6 (e), Fig. 7 and Fig. 8. These plots show the results of the total fracture probability distribution, the age influence and the gender influence, respectively. For comparison, the user is also asked to score the conventional fracture heat map method (using four anatomical views) with the Likert scale table. The total fracture probability distribution using the conventional method is shown in Fig. 6 (a)-(d). The visualization of age and gender influences using the four anatomical views are provided as supplementary material in the appendix. A total of 15 orthopaedic surgeons participated the survey. The average scores for the conventional method and the newly proposed method are shown in Table 3.

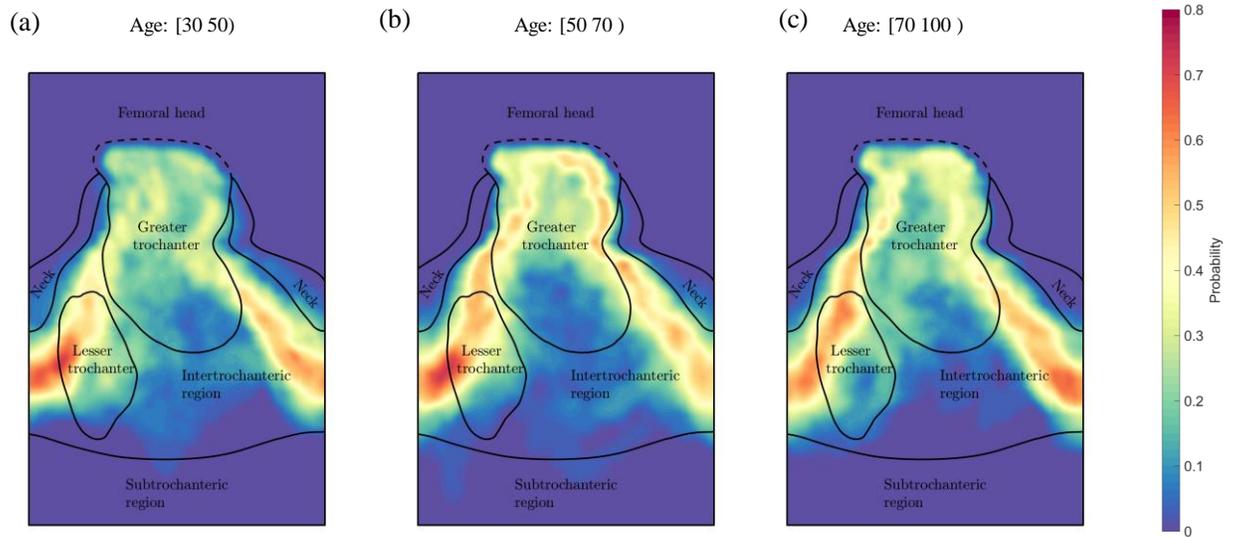

Fig. 7 Proximal femur fracture distribution for three age groups (a) [30 50), (b) [50 70) and (c) [70 100) on the 2D parametric fracture heat map.

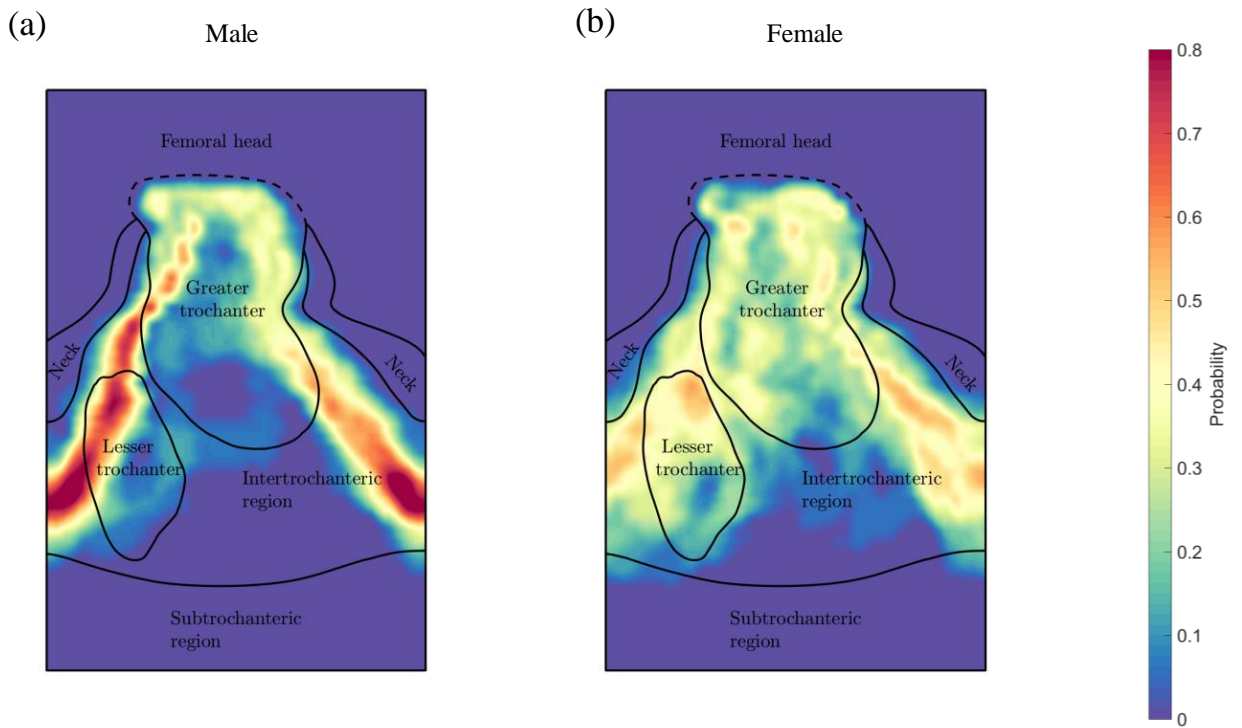

Fig. 8 Proximal femur fracture distribution for (a) male and (c) female over 70 years old on the 2D parametric fracture heat map.

Table 2 The Likert scale table used for accessing the fracture heat map techniques. (The users rate each question using the five-level scale: 1. Strongly disagree; 2. Disagree; 3. Neutral; 4. Agree; 5. Strongly agree.)

| No. | Likert items |
|---|---|
| 1 | The method requires a steep learning curve. |
| 2 | The method has the ability to reflect fracture pattern and frequency faithfully. |
| 3 | The method is efficient for accessing IT fracture characteristics. |
| 4 | The method is compact and information-rich in presenting fracture pattern. |
| 5 | The method is suitable for comparing IT fracture trends, which can be affected by different factors (gender, age, location etc.). |

Table 3 The average scores for the five questions listed in the Likert scale table in accessing the 2D parametric map and conventional method for IT fracture visualization.

|  | 1 | 2 | 3 | 4 | 5 |
|---|---|---|---|---|---|
| **2D parametric map** | 3.69 | 3.62 | 4.13 | 4.31 | 4.06 |
| **Conventional method** | 2.69 | 3.75 | 3.69 | 3.43 | 3.12 |

## 4 Discussions

The fracture probability heat map on the parametric plane can display the full bone surface information with one single, structurally connected heat map. In conventional anatomical view, it requires four images to fully display the fracture frequency information. The neck region in the proximal femur, which is partially covered by lesser and greater trochanter in four anatomical views is more clearly visualized in the 2D parametric fracture map. The anterior and posterior view can be integrated and visualized clearly in one plot using the 2D parametric fracture map. The high-risk fracture region lies most along the intertrochanteric crest and intertrochanteric line. This trend can be determined directly on the 2D parametric map without rotation of the 3D mesh model. The high-risk regions are related to the obliquity fracture, which is the most frequent type of IT fracture. Compared to these high-risk regions, the distal lateral wall and the subtrochanteric region have much lower fracture probability. These regions are mostly associated with reverse obliquity fracture, in which the fracture line usually runs from distal-lateral to proximal-medial. The fracture heat map confirms the rarity of reverse obliquity fracture among IT fractures and the observation is consistent with the previous reports in the literature [18,19].

This study also considers the influence of age and gender on the frequency and the pattern of IT fractures. This study includes the IT fractures cases with the patient's age range from 32 to 97 years old. The osteoporosis is closely related to the age [20,21]. And it can change the strength of the bone. The male to female ratio of IT fracture reported in this study is 1:1.38. The higher incidence of female IT fracture is also reported previously in the literature [22–25]. These evidence indicate that the IT fractures may not be viewed as a unitary case. Considering these factors, the 2D parametric fracture map is then plotted with different age and gender groups. This helps to identify and understand the heterogeneity of IT fracture in these groups.

The influence of the age is shown in Fig. 7. According to the literature study [26–28], the chance of falling onto the greater trochanter increases with age due to the loss of agility and multitasking ability. This explains the increased fracture probability in the greater trochanter region in the older groups observed in Fig. 7. Osteoporosis among the elders can also be a factor, which increases the fragility of bone during falling [29]. Comparing the fracture risk among three age groups, all three share an overall similar fracture distribution pattern in the plots though probability does change locally. During an impact, the compressive and tensile stress distributions depend on the bone structure and dimension. These stress distributions and their maxim values largely determine the fracture pattern, frequency and location during the IT fracture [30–32]. The osteoporosis and increasing probability of falling with aging can increase the probability of IT fracture and the local fracture distribution, but do not alter the overall IT fracture pattern since the bone structure does not change considerably over age.

For the comparison of gender influence, the fracture probability is compared between male and female with the age of 70-year or older. As reported in the previous study, the female generally has a more severe loss of bone with aging compared to male [24]. This reduces the strength of bone due to osteoporosis. When falling onto the greater trochanter, the chance of IT fracture could increase compared with transmitting the force to the femoral neck. In addition, osteoporosis is largely the consequence of trabecular bone loss. This can shift the force to the base of the femoral neck or the adjacent intertrochanteric region [30]. These two factors are considered to be the main causes of the more scattered distribution of IT fracture in the female group over 70 years old.

As to the user experience, the 2D parametric map method outperformed the conventional method in term of compactness and efficiency in visualizing IT fracture pattern and characteristics. It is more suitable for comparing the IT fracture trends among different age and gender groups. The cost of achieving these advantages is a steeper learning curve (3.69 vs. 2.69) to relate the regions in the 2D parametric map to the actual 3D model. In terms of presenting information faithfully, both methods have similar scores (3.62 vs. 3.75).

## 5 Conclusions

In this study, 100 cases of IT fractures are visualized and analyzed using the 2D parametric fracture probability heat map. The proposed map projection technique can project the high dimensional proximal femur fracture information onto a single 2D plane. This technique retains the original 3D proximal femur structure information and can be used to visualize anterior and posterior views simultaneously. This is more convenient for IT fracture visualization compared to the anatomical view representation. In addition, the new technique allows for a straightforward calculation and visualization of the statistics of a large dataset. The high IT fracture risk region of studied 100 patients is identified and visualized using the 2D parametric fracture heat map technique. The result shows that the intertrochanteric line and crest region are the high-risk regions. High fracture probability appeared in these regions is consistent with the biomechanical stress analysis reported in the literature.

The study also shows that the IT fracture pattern is related to age and gender closely. With the increase of age, the high probability of falling on greater trochanter and osteoporosis increase

IT fracture probability, but spread out the fracture distribution. The greater loss of bone in female results in a more diffused distribution of IT fracture. Therefore, this study suggests that in clinical diagnosis and treatment of IT fracture, close attention should be paid to the high risk identified region in this study. One should take into the consideration the gender and age of the patient since the fracture probability and distribution shows strong relations with these factors.

## 6 Declarations

The authors would like to thank the Institutional Ethics Committee of PuRen Hospital for providing the data used in this study. One of the authors, Rong Liu would like to acknowledge the support from Wuhan City Health and Family Planning Scientific Research Project (Grant No. WX16B21), Hubei Province Health and Family Planning Scientific Research Project (Grant No. WJ2017F032, No. WJ2018H0042) and Metallurgical Safety and Health Branch of China Metals Society Health Research Project (Grant No. JKWS201620).

# Appendix A. Supplementary material

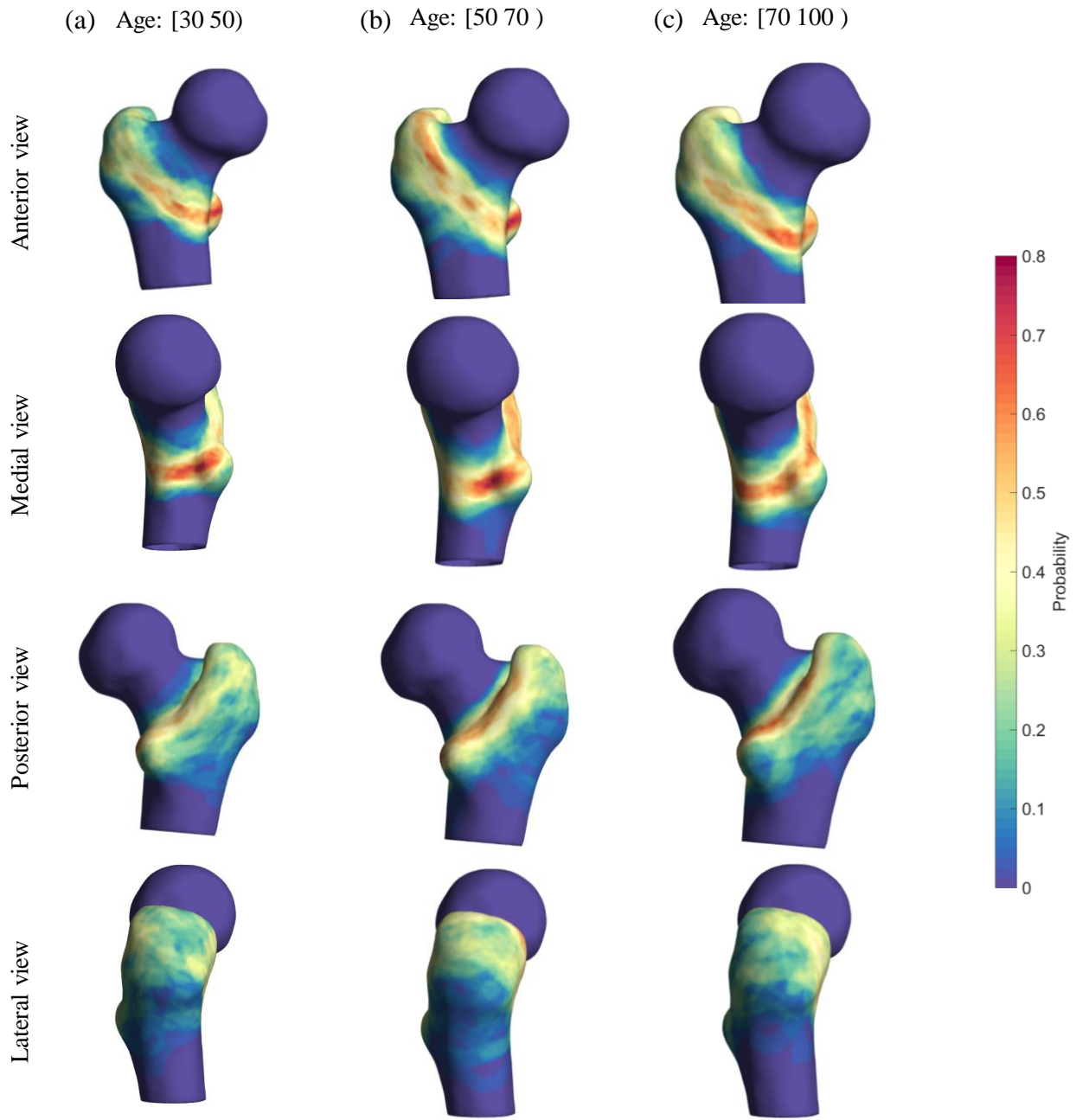

Fig. A.1 Proximal femur fracture distribution for three age groups (a) [30 50), (b) [50 70) and (c) [70 100) using the four anatomical views.

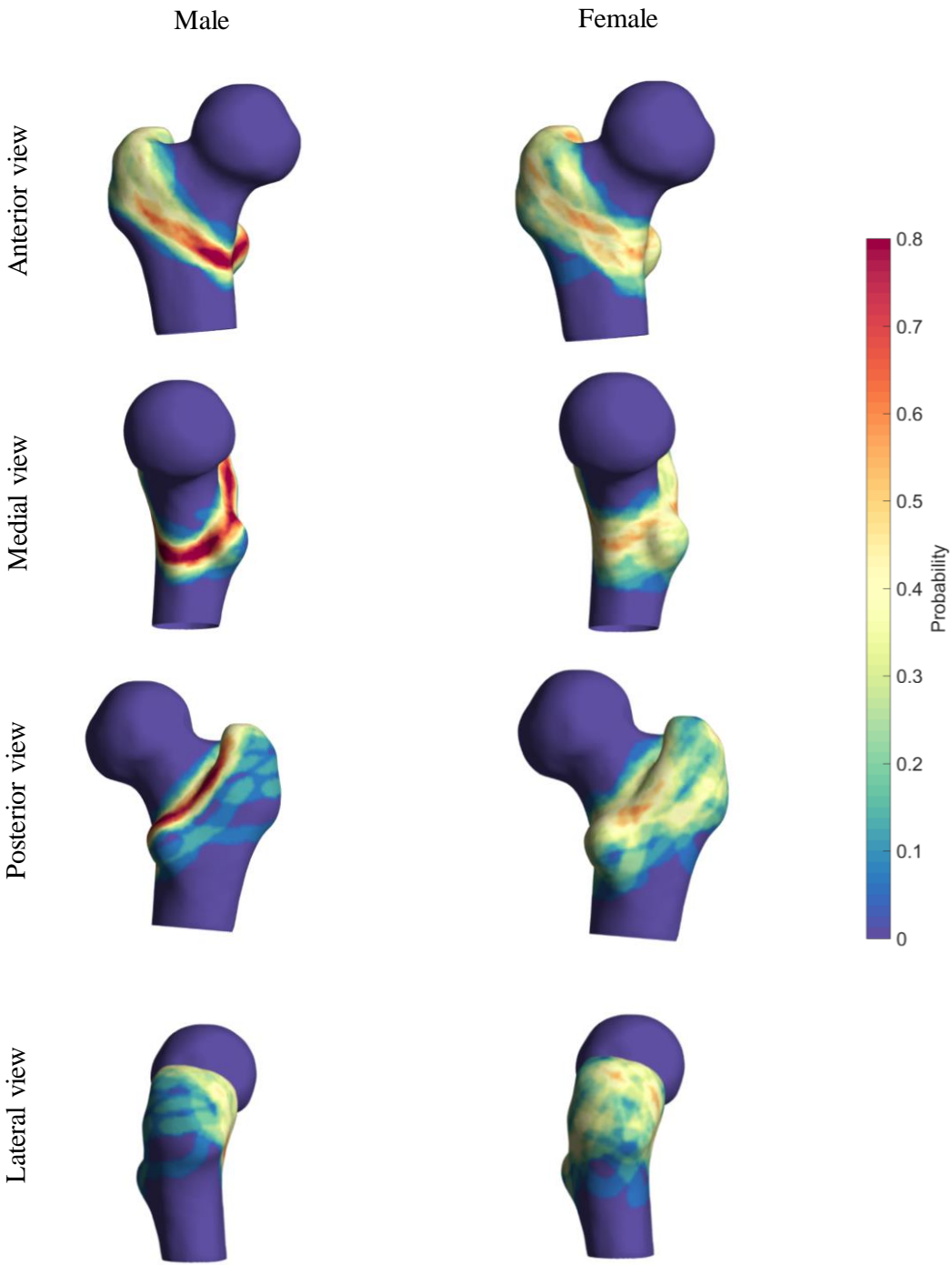

Fig. A.2 Proximal femur fracture distribution for two gender groups (a) male, (b) female using the four anatomical views.